\documentclass[aip, amsmath,amssymb, reprint]{revtex4-2}

\usepackage{graphicx}
\usepackage{dcolumn}
\usepackage{bm}
\usepackage[utf8]{inputenc}
\usepackage[T1]{fontenc}
\usepackage{mathptmx}
\usepackage{etoolbox}
\usepackage{color}

\newcommand{\ds}{\displaystyle}
\begin{document}

\title[Inverse spin galvanic effect in proximitized superconductor/paramagnet systems]{%
Inverse spin galvanic effect in proximitized superconductor/paramagnet systems}
\author{S. V. Mironov}
\affiliation{Institute for Physics of Microstructures, Russian Academy of Sciences, 603950 Nizhny Novgorod, GSP-105, Russia}
\author{A. S. Mel'nikov}
\affiliation{Institute for Physics of Microstructures, Russian Academy of Sciences, 603950 Nizhny Novgorod, GSP-105, Russia}
\affiliation{Moscow Institute of Physics and Technology (National Research University), Dolgoprudnyi, Moscow region, 141701 Russia}
\author{A. I. Buzdin}
\email[Author to whom correspondence should be addressed: ]{alexandre.bouzdine@u-bordeaux.fr}
\affiliation{University Bordeaux, LOMA UMR-CNRS 5798, F-33405 Talence Cedex, France}
\affiliation{World-Class Research Center ``Digital biodesign and personalized healthcare'', Sechenov First Moscow State Medical University, Moscow 119991, Russia}

\date{\today }

\begin{abstract}
We show that the interplay between spin-orbit coupling (SOC) and the paramagnetic response of itinerant electrons in proximitized superconductor/paramagnet systems gives rise to the inverse spin galvanic effect, i.e. generation of magnetic moment under the influence of the charge current. Depending on the sign of the SOC constant and the system temperature, the corresponding contribution to the  magnetic response of the superconductor can be either diamagnetic or paramagnetic. We discuss the relevance between the discovered phenomena and the recent experiments on Pt/Nb heterostructures as well as the puzzling sign change of the magnetic response observed  in clean Ag coated Nb cylinders.
\end{abstract}

\maketitle

\preprint{AIP/123-QED}

The interplay between charge currents and spin polarization in
superconducting systems with broken inversion symmetry stays in the focus of
intensive research for almost three decades.\cite{Mineev_Review,
Agterberg_review} From a symmetry standpoint, the absence of inversion
symmetry in the direction along a certain unit vector ${\vec{n}}$ results
in the appearance of contribution $\propto {\vec{p}}\cdot \left( {\vec{n}}%
\times {\vec{\sigma}}\right) $ to the system energy, and in the ground state
the electron momentum ${\vec{p}}$ becomes coupled to its spin ${\vec{\sigma}}
$ forming the phase-modulated helical state.\cite{Agterberg_review,
Edelstein_HelicalStates, Mineev_HelicalStates} Among the great variety of
physical consequences coming from this Rashba spin-orbit coupling
(SOC) in superconductors one may recognize the specific magneto-electric
phenomena revealing themselves through the generation of spontaneous
electric currents parallel to the vector ${\vec{n}}\times {\vec{h}}$ in
superconducting systems where electron spins are polarized by the exchange
field ${\vec{h}}$ of the adjacent ferromagnet.\cite{Mironov-PRL-17,
Devizorova-PRB-21, Putilov-PRB-24} Similar phenomena arise in the Josephson
phase batteries ($\varphi _{0}$ junctions) where the combination of the
exchange field and SOC induces the spontaneous phase difference across the
junction.\cite{Buzdin_Phi, Krive, Reynoso, Kouwenhoven}

The presence of the term $\propto {\vec p}\cdot \left({\vec n}\times {\vec
\sigma}\right)$ in the system energy naturally suggests the existence of the
inverse magneto-electric phenomena, namely, the generation of spin polarized
states with the spin orientation parallel to the vector ${\vec n}\times {%
\vec j}$ under the influence of the transport charge current ${\vec j}$.
Previously, similar phenomena has been extensively studied in semiconductors
with strong SOC where the charge currents were shown to induce both the spin
currents (so-called spin Hall effect)\cite{SpinHall_rev} and nonzero spin
polarization (this phenomenon is often called as inverse spin
galvanic effect, Edelstein effect or just current induced spin
polarization)\cite{Ivchenko, Edelstein_1990, Aronov, Golub, Ganichev_rev, Ivchenko_book}. Up to now
several models were proposed to uncover the possible microscopic mechanisms
responsible for the inverse spin galvanic effect in superconductors. First,
the formation of the helical states directly influences the spin structure
of superconducting pair correlations carrying charge current which should
result in subsequent generation of magnetic moment for both clean and dirty
superconductors.\cite{Edelstein_1, Edelstein_2} Similarly, in Josephson
systems the dc current flowing through the $\varphi_0$ junction was shown to
induce both equilibrium magnetic moment\cite{Konschelle_Bergeret} and the
effective field which acts on electron spins and gives rise to the
magnetization precession.\cite{Konschelle, Shukrinov_rev} Another mechanism
of the inverse spin galvanic effect relies on the current induced generation
of the long-range spin-triplet correlations beyond the quasiclassical
approximation.\cite{Bobkova-PRB-2017, Silaev} Recently this approach was
applied to explain the puzzling paramagnetic contribution to the magnetic
response observed in Pt/Nb structures.\cite{Lee_exp} 

%Note that all mentioned theoretical models describing spin galvanic phenomena are based on the analysis of microscopic details of the Cooper pair wave function in the presence of spin-orbit coupling.

In this Letter, stimulated by the experiments [\onlinecite{Lee_exp}] we suggest an alternative mechanism of the inverse spin galvanic effect in superconducting systems coming from the interplay between spin orbit coupling and paramagnetic response of itinerant electrons (note that the Pt layer in [\onlinecite{Lee_exp}] is a Stoner enhanced paramagnet). According to this mechanism, the superconducting Meissner current screening the external magnetic field can induce the magnetization ${\bf M}$ inside the paramagnetic (PM) layer due to SOC. This magnetization, in its turn, renormalizes the magnetic field and produces the correction to the Meissner currents, which results in the additional quadratic over the SOC constant contribution to the magnetization and the magnetic field inside the sample. The appearance of the described first and second-order corrections to the magnetic field is the direct consequence of the inverse spin galvanic effect.

\begin{figure}[t!]
\begin{center}
\includegraphics[width=1\linewidth]{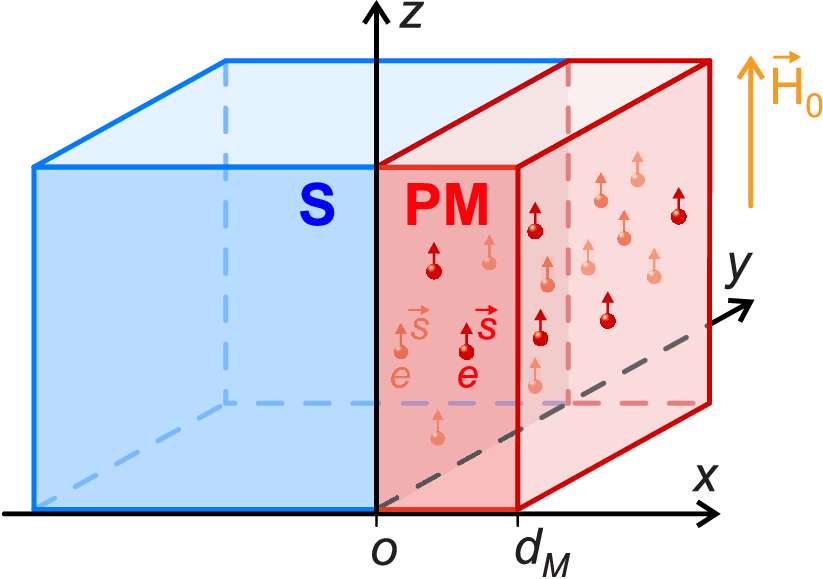}
\end{center}
\caption{Sketch of the superconductor/paramagnet heterostructure in an external magnetic field ${\bf H}_0$.}\label{Fig1}
\end{figure}

To elucidate our main idea we consider a layered hybrid system consisting of
a paramagnetic material of the thickness $d_M$ positioned on top of a
thick superconductor (S). The sketch of the system is shown in Fig.~\ref{Fig1}. Keeping in mind the setup used in the experiment of Ref.~[\onlinecite{Lee_exp}] we consider the situation when the sample is put in
the external magnetic field $\mathbf{H}_0$ directed along the $z$ axis. We
assume that the superconductivity is well-developed and the interface
between S and PM layers is transparent for electrons so that Cooper pairs
freely penetrate from superconductor to paramagnet and induce the
superconducting correlations there. Considering the dirty limit we assume
the local relation between the superconducting current $\mathbf{j}_s$ and
the vector potential $\mathbf{A}$. In the bulk of the superconductor this relation takes the London form $\mathbf{j}%
_s=-(c/4\pi)\lambda^{-2}\mathbf{A}$ where $\lambda$ is the London
penetration depth and $c$ is the speed of light. For simplicity we restrict ourselves to the limit
 when the PM layer thickness $d_M$ is smaller compared to the
superconducting correlation length $\xi_n$, which determines the typical
decay scale for  superconducting correlations inside the
non-superconducting material in the dirty limit. The latter condition allows
us to neglect the spatial variation of the Cooper pairs density across the
S/PM structure and consider the length $\lambda$ to be constant throughout
the system. Finally, we assume that the paramagnetic material has a strong
Rashba spin-orbit coupling.

The system thermodynamic potential relevant for the case of the fixed
external magnetic field and temperature can be written in the form $F=S\int_{-\infty
}^{d_{M}}f(x)dx$ with 
\begin{equation}
f(x)=\frac{\left( B_{z}-H_{0}\right) ^{2}}{8\pi }+\frac{A_{y}^{2}}{8\pi
\lambda ^{2}}+\left( \alpha M_{iz}A_{y}+\frac{\beta }{2}M_{iz}^{2}-M_{iz}B_{z}%
\right) ,  \label{F_def}
\end{equation}%
where $x$ is the coordinate across the layers, $S$ is the surface area of the S/PM interface, $%
\alpha $ is the constant describing the strength of SOC in the PM layer, $\mathbf{B}=\mathrm{curl~}\mathbf{A}$ is the magnetic field, ${\bf M}_i$ is the part of magnetization corresponding to the spin polarization of itinerant electrons inside the paramagnetic layer while the total magnetization ${\bf M}=\left({\bf B}-{\bf H}_0\right)/4\pi$ contains an additional contribution ${\bf M}_s$ coming from the superconducting currents so that ${\bf M}={\bf M}_s+{\bf M}_i$. The constant $\beta=4\pi+1/\chi$ is determined by the magnetic susceptibility $\chi$ of itinerant electrons in the normal state when ${\bf M}_s=0$ so that ${\bf M}_i=\chi {\bf H_0}$. Note that the condition $\chi >0$ reflecting the stability of the paramagnetic state means that $\beta>4\pi$. The last three terms in Eq.~(\ref{F_def})
describe the paramagnetic material and its interaction with magnetic field
and current, so they contribute only for $0<x<d_M$. The origin of the unusual, linear over $A_{y}$
term is related to the Rashba spin-orbit interaction in the thin
paramagnetic layer which generates $\sim \Psi ^{\ast }\left( {\bf n}
\times {\bf h}\right) \left( -i \nabla +2\pi{\bf A}/\Phi_0\right) \Psi $ invariant in the Ginzburg-Landau free energy with $\Psi$ being the superconducting order parameter and $\Phi_0$ being the magnetic flux quantum.\cite{Samokhin_2004, Kaur_2005} It is this anomalous linear over $A_{y}$ term describing the direct
mechanism of the magnetization generation by the superconducting current
(proportional to ${\bf A}$) which is responsible for the inverse spin galvanic effect. We assume here that the
exchange field ${\bf h}$ is produced by the polarization of the
paramagnet, neglecting small contribution from Zeeman effect. We may
evaluate the coefficient 
\begin{equation}\label{alpha_est}
\alpha \sim \alpha _{so}\frac{\Phi _{0}}{M_{0}\xi
_{0}\lambda ^{2}},
\end{equation}
where $\alpha _{so}=\left(v_{so}/v_{F}\right)\left(h_{0}/T_{c}\right)$, $v_{s0}$ is the spin-orbit velocity, $v_F$ is the Fermi velocity, $h_{0}$ and $M_{0}$ are the saturated exchange field and magnetic moment, respectively, $T_c$ is the critical temperature of the superconducting transition, and $\xi_0$ is the superconducting coherence length. Remarkably, from Eq.~(\ref{alpha_est}) one sees that the value $\alpha$ strongly depends on temperature $T$ near the critical temperature $T_c$ so that $\alpha\propto\lambda^{-2}\propto\left|\Psi\right|^2\propto\left(T_c-T\right)$. As a result, the phenomena associated with SOC have peculiar temperature dependencies which, as we will show below, should allow their identification among other types of similar effects.

%$\frac{1}{\lambda ^{2}}\approx \left\vert \Psi \right\vert ^{2}\xi _{0}^{2}e^{2},$ $\alpha M_{z}\sim \frac{ v_{so}}{v_{F}}\left\vert \Psi \right\vert ^{2}\xi _{0}e\frac{h}{T_{c}}\sim  \frac{v_{so}}{v_{F}}\left\vert \Psi \right\vert ^{2}\xi _{0}e\frac{h_{0}}{ T_{c}}\frac{M_{z}}{M_{0}}\sim \alpha _{so}\frac{M_{z}}{M_{0}}\frac{1}{ \lambda ^{2}}\frac{\Phi _{0}}{e\xi _{0}}\sim \alpha _{so}\frac{\Phi _{0}}{ M_{0}\xi _{0}\lambda ^{2}}M_{z}$

The thermodynamic potential (\ref{F_def}), as a functional, is determined by two independent functions, namely, $A_y(x)$ and $M_{iz}(x)$. Varying the potential $F$ with respect to $M_{iz}$ we find the equation 
\begin{equation}
\alpha A_{y}+\beta M_{iz}-B_{z}=0,
\end{equation}%
which gives 
\begin{equation}
M_{iz}=\frac{1}{\beta }B_{z}-\frac{\alpha }{\beta }A_{y}.  \label{Eq_M}
\end{equation}%
In Eq.~(\ref{Eq_M}) the first term stands for the usual paramagnetic
response inside the PM layer while the second one arises due to SOC. It is this second term which is responsible for the inverse spin galvanic effect, i.e. generation of magnetization under the influence of the superconducting current proportional to $A_y$. Since
the constant $\alpha $ can have arbitrary sign, the corresponding
contribution to the local magnetic moment can be either paramagnetic or
diamagnetic depending on the peculiarities of SOC in the specific material. In addition, the dependence of the second term in Eq.~(\ref{Eq_M}) on temperature differs from the one for the first term, which makes it possible to distinguish between the usual paramagnetism and SOC induced inverse spin galvanic effect in experiments.

Varying the thermodynamic potential with respect to $A_y$ and accounting
that $B_z=\partial A_y/\partial x$ we obtain: 
\begin{equation}
-\frac{\partial^2 A_y}{\partial x^2}+\frac{A_y}{\lambda^2}+4\pi\alpha M_{iz} +
4\pi\frac{\partial M_{iz}}{\partial x}=0.
\end{equation}
Substituting the obtained expression for $M_{iz}$ and performing algebraic
simplifications we find that inside the paramagnetic layer the vector
potential satisfies the London equation 
\begin{equation}  \label{Eq_A}
\frac{\partial^2 A_y}{\partial x^2}-\frac{1}{\lambda_M^2} A_y =0,
\end{equation}
where the London penetration depth characterizing the decay scale of the
magnetic field becomes renormalized: 
\begin{equation}\label{Lambda}
\frac{1}{\lambda_M^2}= \left(\frac{1}{\lambda^2} -\frac{4\pi\alpha^2}{\beta}%
\right)\frac{\beta}{\beta-4\pi}.
\end{equation}
We may see that the SOC leads to the increase of the effective London penetration depth inside the paramagnet. For very strong SOC the expression (\ref{Lambda}) for $\lambda_M^{-2}$ may become negative. In fact, it means that at $4\pi\alpha^2/\beta>\lambda^{-2}$ the system should generate the magnetic moment even in the absence of the applied external magnetic field. So the soft paramagnetic system with strong SOC is unstable towards a spontaneous magnetic transition.

The solution of the London equation (\ref{Eq_A}) inside the superconductor
(for $x<0$) reads: 
\begin{equation}
A_{y}=A_{0}\exp \left( \frac{x}{\lambda }\right) ,  \label{A_Gen_1}
\end{equation}%
while inside the paramagnetic layer one finds 
\begin{equation}
A_{y}=A_{1}\cosh \left( \frac{x}{\lambda _{M}}\right) +A_{2}\sinh \left( 
\frac{x}{\lambda _{M}}\right) .  \label{A_Gen_3}
\end{equation}%
The constants $A_{0}$, $A_{1}$ and $A_{2}$ should be found from the boundary
conditions which require the continuity of the vector potential at $x=0$ and
the continuity of the combination $\left(B_{z}-4\pi M_{iz}\right)$ both
at the S/PM interface and the outer boundary of the PM layer (outside the
sample $B_{z}=H_{0}$). The latter condition follows from the continuity of the magnetic field strength $H_0=B_z-4\pi \left(M_{iz}+M_{sz}\right)$ as well as the continuity of the magnetization $M_{sz}$ associated with the superconducting currents. Note that since the magnetization arising due to the spin-polarization of itinerant electrons is nonzero only inside the PM layer and the interfaces are assumed to have atomic scale, the magnetization reveals jumps at the boundaries of PM layer which should result in the step-like behavior of the magnetic field $B_z$ near the planes $x=0$ and $x=d_M$. Taking into account Eq.~(\ref{Eq_M}) and solving the
resulting systems of 3 equations for $A_{0}$, $A_{1}$ and $A_{2}$ in the
limit $d_{M}\ll \lambda $ (see supplemental material for the calculation
details) we obtain that inside the superconductor the magnetic field takes
the form 
\begin{equation}
B_{z}(x)=\left( B_{0}+\Delta B\right) \exp \left( \frac{x}{\lambda }\right) ,
\label{B_res}
\end{equation}%
where $B_{0}=H_{0}\left( 1-d_{M}/\lambda \right) $ is the magnetic field at
the S/PM interface (at $x=0$) in the absence of SOC while the value 
\begin{equation}
\Delta B=\frac{4\pi }{(\beta -4\pi)}\left( \alpha^2 \lambda
-\alpha\right)d_{M} H_{0}
\end{equation}%
is the correction to the magnetic field coming from the inverse spin
galvanic effect. The first quadratic over $\alpha $ term in the brackets
always provides a paramagnetic contribution to the screening while the
contribution from the second linear over over $\alpha $ term may
be either negative or positive, depending on the sign of the spin-orbit
interaction constant. The sign of this constant depends on the details of
the electronic structure and may vary depending on the choice of material. 
Interestingly that for negative $\alpha$ we always obtain a
paramagnetic contribution due to the spin-orbit effect.  We believe that
such a situation is realized in Pt/Nb structures,\cite{Lee_exp}
which provides a possible explanation of the observed effect. Note that the paramagnetic contributions to the value $\Delta B$ have a very different $\propto\left(T_c-T\right)$ and $\propto\left(T_c-T\right)^{3/2}$ temperature dependencies compared to the standard $\propto\left(T_c-T\right)^{1/2}$ diamagnetic contribution in $B_0$. This means that just below $T_c$ the diamagnetism should dominate, but at lower temperatures the paramagnetic contribution may strongly attenuate it.

\begin{figure}[t!]
\begin{center}
\includegraphics[width=0.95\linewidth]{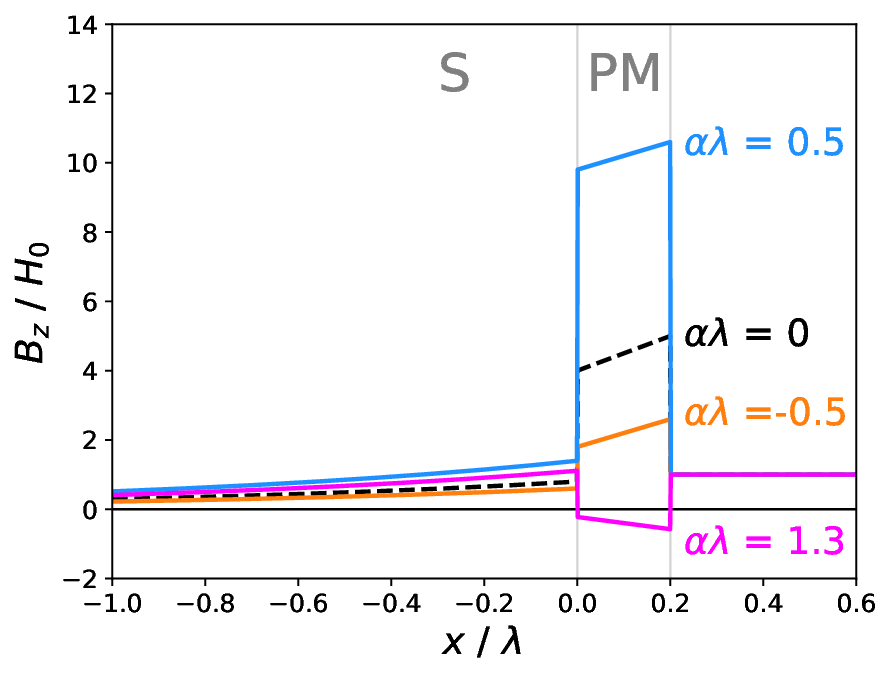}
\end{center}
\caption{Spatial profiles of the magnetic field inside the superconductor/paramagnet bilayer. In our calculation we took $\beta=5\pi$, $d_M=0.2\lambda$ and four different values of the dimensionless parameter $\alpha\lambda$ shown near the corresponding curves.}\label{Fig2}
\end{figure}

For positive $\alpha $ the situation depends on the ratio
between the lengthscales $\alpha^{-1}$ and $\lambda $: for $\alpha \lambda
<1$ the spin-orbit contribution will always generate diamagnetic
response, while for  $\alpha \lambda >1$ the response should be
paramagnetic. Taking into account that $\alpha \propto \left( T_{c}-T\right) $ where $T$ is the system temperature we see
that $\alpha \lambda \propto \left( T_{c}-T\right) ^{1/2}$ and just
below $T_{c}$ the response should be diamagnetic, while at lower
temperature it may switch to the paramagnetic one. Note that very similar
behavior was observed in the experiments [\onlinecite{Visani}] on clean Ag coated Nb
proximity cylinders. We may speculate that  the discussed spin-orbit
mechanism may be at the origin of the puzzling results [\onlinecite{Visani}].  

\begin{figure}[t!]
\begin{center}
\includegraphics[width=0.95\linewidth]{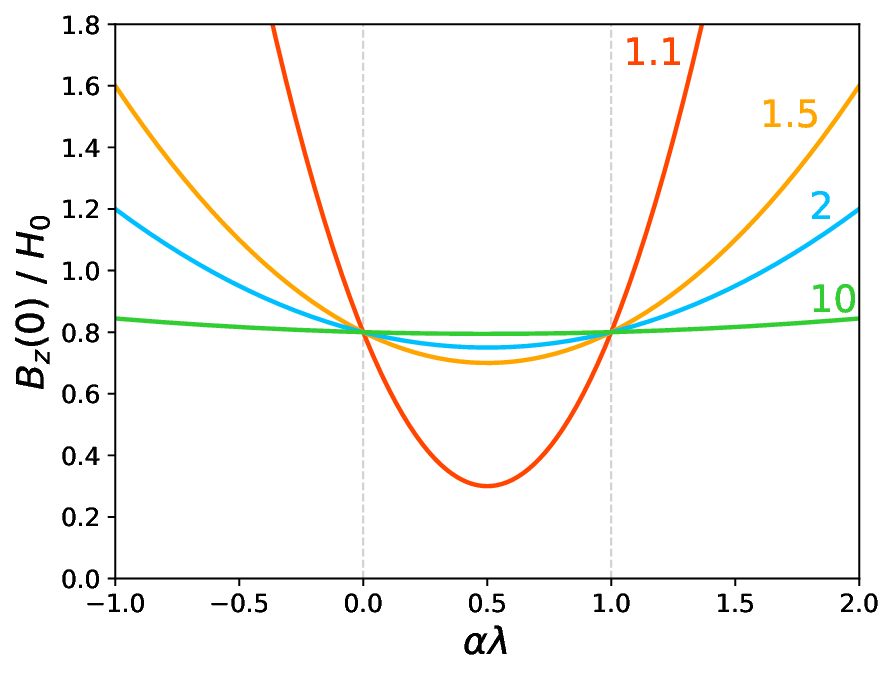}
\end{center}
\caption{Dependencies of the magnetic field $B_z(0)$ inside the superconductor at the superconductor/paramagnet interface on the parameter $\alpha\lambda$ for the following different values of the parameter $\beta/4\pi$: $1.1$, $1.5$, $2$ and $10$ (the values are shown near the corresponding curves). In our calculations we took $d_M=0.2\lambda$.}\label{Fig3}
\end{figure}

In order to estimate experimental achievability of the regime $\alpha\lambda>1$ we consider Eq.~(\ref{alpha_est}) and take into account that $\Phi_0/(\xi_0\lambda)\sim H_{cm}/\sqrt{1-(T/T_c)}$ where $H_{cm}$ is the thermodynamic critical magnetic field of the superconductor. The saturated magnetization $M_0$ may approach the value $\sim n\mu _{B}$ (here $n$ is the electron density and $\mu_B$ is the Bohr magneton) which is comparable to the critical field $H_{cm}$. In the case of the strong spin-orbit coupling the parameter $\alpha_{so}$ may become larger than $1$ due to the large ratio $(h_{0}/T_c)$. These estimates show that the crossover from the diamagnetic to the paramagnetic response with the decrease of temperature seems to be possible. The typical profiles of the magnetic field inside the heterostructure for $\alpha>0$ and $\alpha<0$ are shown in Fig.~\ref{Fig2}. To illustrate the possible crossovers between paramagnetic and diamagnetic response of the magnetic field inside the superconductor in Fig.~\ref{Fig3} we also plot several dependencies of the magnetic field $B_z(0)$ inside the superconductor at the S/PM interface as a function of the parameter $\alpha\lambda$ for different values of $\beta$. 

The parameter $\alpha\lambda$  affects also the qualitative behavior of the magnetization inside the paramagnet. The spatial profile of this magnetization for $d_M\ll\lambda_M$ reads (see supplemental material for calculation details):
\begin{equation}\label{Eq_M_res}
M_{iz}=H_0\frac{\left(1-\alpha\lambda\right)}{\beta-4\pi}\left[1+\frac{x-d_M}{\lambda}-\frac{4\pi \left(1-\alpha\lambda\right)}{\left(\beta-4\pi\right)}\alpha d_M \right].
\end{equation}
One sees that for $\alpha\lambda>1$ the magnetization direction inside the PM layer is opposite to the one realized in the absence of the SOC. Thus, varying the temperature in the vicinity of $T_c$ one may provoke the reversal of magnetization.

An alternative way to perform controllable switching between between paramagnetic and diamagnetic response of the superconductor can be realized if one adds an insulator (I) on top of the PM layer and another normal metal layer (N) on top of the insulator to form a S/PM/I/N heterostructure (from the bottom to the top layer). In this case, by applying a voltage between S and N layers, it should be possible to change the sign of the constant $\alpha$ in the PM and switch the Meissner response from diamagnetic to paramagnetic at a fixed temperature (we are grateful to the reviewer of our paper who has drawn our attention to this opportunity).

Note that in our previous works\cite{Mironov_2018, Devizorova_2019} we discussed the
electromagnetic proximity effect in superconductor (S)/ferromagnet (F) bilayers, where the magnetic field
in superconductor was induced by the supercurrent screening the F layer
magnetic moment. Similar physics works in the situation considered above 
but with a magnetic moment induced by the Meissner current due to the
applied magnetic field. 

In experiment\cite{Lee_exp} the paramagnetic layer has the thickness of several
tenths of ${\rm nm}$ and the region with a Rashba spin-orbit interaction is smaller
than the total thickness. In the framework of our model we can describe this situation by
adding an additional superconducting layer with a thickness $d_{S}$ on top of the paramagnet. We assume that the superconducting
correlations are present in this layer with $d_{S}\ll\lambda$ and repeating the previous calculations, we retrieve the expression
(\ref{B_res}) with $B_{0}=H_{0}\left[1-(d_M+d_{s})/\lambda\right]$. This is quite
natural because the thin superconducting layer provides just additional screening
of the applied field.

%In conclusion, we have revealed a new type of proximity effect between superconductor and paramagnetic metal with a strong spin-orbit interaction. It may be considered as an inverse spin galvanic effect and is related to the magnetic polarization generation by a screening current. 

In conclusion, we have shown that the proximity effect between superconductor and paramagnetic metal with a strong spin-orbit interaction gives rise to an additional phenomenon, namely, inverse spin galvanic effect which is related to the magnetic polarization generation by a screening current. The obtained results provide some hints to the understanding of unusual proximity effects observed in Pt/Nb structures [\onlinecite{Lee_exp}] and Ag coated Nb cylinders [\onlinecite{Visani}]. The predicted special temperature dependence of the magnetic response may help to identify this type of the proximity effect.

\bigskip

\section*{Supplementary Material}

Detailed calculations of the magnetization distribution and the magnetic field inside the S/PM/S system.

\bigskip 
\begin{acknowledgments}
This work was supported by the Russian Science Foundation (Grant No. 20-12-00053) in part related to the calculation of the magnetic field profiles in the superconductor/paramagnet bilayer and Ministry of Science and Higher Education of the Russian Federation within the framework of state funding for the creation and development of World-Class Research Center (WCRC) ``Digital biodesign and personalized healthcare'', grant no. 075-15-2022-304, in part related to the analysis of temperature dependencies of the magnetic field amplitude. S. V. M. acknowledges the financial support of the Foundation for the Advancement of Theoretical Physics and Mathematics BASIS (Grant No. 23-1-2-32-1). 
\end{acknowledgments}

\bigskip

\section*{Data Availability Statement}

The data that support the findings of this study are available from the
corresponding author upon reasonable request.

\renewcommand{\theequation}{S\arabic{equation}}

\setcounter{equation}{0}

\section*{Supplementary material}

Here we calculate the distribution of the magnetization and the magnetic field inside the system consisting of superconducting half-space $x<0$, paramagnetic layer with spin-orbit coupling occupying the region $0<x<d_M$ and the second superconducting layer of the finite thickness $d_s$ positioned at $d_M<x<d_M+d_s$. The case $d_s=0$ corresponds to the superconductor/paramagnet bilayer.

The solution of the London equation inside the superconducting half-space takes the form
\begin{equation}\label{A_S1}
A_{y}=A_{0}\exp \left( \frac{x}{\lambda }\right).  
\end{equation}
Inside the paramagnetic layer the vector potential reads 
\begin{equation}\label{A_PM}
A_{y}=A_{1}\cosh \left( \frac{x}{\lambda _{M}}\right) +A_{2}\sinh \left( 
\frac{x}{\lambda _{M}}\right),  
\end{equation}
while in the second superconducting layer
\begin{equation}\label{A_S2}
A_{y}=A_{3}\cosh \left( \frac{x-d_M-d_s}{\lambda}\right) +A_{4}\sinh \left( 
\frac{x-d_M-d_s}{\lambda}\right).  
\end{equation}
The constants $A_{0}$, $A_{1}$, $A_{2}$, $A_{3}$ and $A_{4}$ should be found from the boundary conditions. To simplify the calculations we assume that the thicknesses of both paramagnetic and second superconducting layers are much smaller than $\lambda$. So, in what follows we will consider only the terms which are linear over these thicknesses neglecting the higher-order contributions.

The continuity of the vector potential at the interfaces $x=0$ and $x=d_M$ gives the first two equations:
\begin{equation}\label{Eq1}
A_0=A_1,  
\end{equation}
\begin{equation}\label{Eq2}
A_1 +A_2\left(d_M/\lambda_M\right)= A_3 -A_4 \left(d_s/\lambda\right).
\end{equation}
The three remaining equations follow from the continuity of the combination $\left(B_z-4\pi M_{iz}\right)$ at $x=0$, $x=d_M$ and $x=d_M+d_s$ (we assume the presence of the magnetic field $H_0$ outside the sample). With the required accuracy the magnetization associated with the spin polarization of itinerant electrons inside the paramagnetic layer has the form
\begin{equation}\label{Eq_M_res}
M_{iz}=-\frac{\alpha}{\beta}\left(A_1+A_2\frac{x}{\lambda_M}\right) + \frac{1}{\beta\lambda_M}\left(A_1\frac{x}{\lambda_M}+A_2\right).
\end{equation}
Then taking into account that $B_z=\partial A_y/\partial x$ we obtain the following set of equations:
\begin{equation}\label{Eq3}
\frac{1}{\lambda}A_0=\frac{1}{\lambda_M}\left(1-\frac{4\pi}{\beta}\right)A_2+\frac{4\pi\alpha}{\beta}A_1,
\end{equation}
\begin{equation}\label{Eq4}
\begin{array}{c}{\ds 
\frac{1}{\lambda_M}\left(1-\frac{4\pi}{\beta}\right)\left(A_1\frac{d_M}{\lambda_M}+A_2\right)+\frac{4\pi\alpha}{\beta}\left(A_1+A_2\frac{d_M}{\lambda_M}\right)=}\\{}\\{\ds =\frac{1}{\lambda}\left(A_4-A_3\frac{d_s}{\lambda}\right),}
\end{array}
\end{equation}
\begin{equation}\label{Eq5}
\frac{1}{\lambda}A_4=H_0.
\end{equation}
Expressing the constant $A_2$ from Eq.~(\ref{Eq3}) and $A_3$ from Eq.~(\ref{Eq2}) and then substituting all the constants except $A_0$ into Eq.~(\ref{Eq4}) after algebraic simplifications we obtain:
\begin{equation}\label{Eq4b}
A_0=H_0\lambda\left[1-\frac{d_M +d_s}{\lambda}+\frac{4\pi }{\beta-4\pi}\left(\alpha^2\lambda-\alpha\right)d_M \right].
\end{equation}

The obtained expression (\ref{Eq4b}) allows one to find the magnetization induced in the paramagnetic layer:
\begin{equation}\label{Eq_M_res}
\frac{M_{iz}}{H_0}=\frac{\left(1-\alpha\lambda\right)}{\beta-4\pi}\left[1+\frac{x-d_M-d_s}{\lambda}-\frac{4\pi \left(1-\alpha\lambda\right)}{\left(\beta-4\pi\right)}\alpha d_M \right].
\end{equation}

Finally, we calculate the magnetic field profile inside the heterostructure. Inside the superconducting half-space the magnetic field reads
\begin{equation}\label{B_res}
B_z=H_0\left[1-\frac{d_M +d_s}{\lambda}-\frac{4\pi \left(1-\alpha\lambda\right)}{\beta-4\pi}\alpha d_M \right]\exp\left(\frac{x}{\lambda}\right).
\end{equation}
Inside the paramagnetic layer it takes the form
\begin{equation}\label{B_PM}
\begin{array}{c}{\ds
B_z=H_0\frac{\beta}{\beta-4\pi}\left\{\left(1 -\frac{4\pi\alpha^2 \lambda^2}{\beta}
\right) \frac{x}{\lambda} +\right.}\\{}\\{\ds \left.+\left(1-\frac{4\pi\alpha\lambda}{\beta}\right)\left[1-\frac{d_M +d_s}{\lambda}+\frac{4\pi }{\beta-4\pi}\left(\alpha^2\lambda-\alpha\right)d_M \right]\right\},}
\end{array}
\end{equation}
while inside the second superconducting layer it reads
\begin{equation}\label{B_S2}
B_z=H_0\left(1+\frac{x-d_M-d_s}{\lambda}\right).  
\end{equation}

\end{document}